\documentclass[%
 reprint,
 amsmath,amssymb,
 aps,
 prl,
superscriptaddress,
citeautoscript
]{revtex4-1}

\usepackage{graphicx}% Include figure files
\usepackage{dcolumn}% Align table columns on decimal point
\usepackage{bm}% bold math
\usepackage{color}
\usepackage{hyperref}% add hypertext capabilities
\usepackage[columnwise,mathlines]{lineno}% Enable numbering of text and display math
\usepackage{siunitx}
\usepackage{chemformula}

\begin{document}

%\preprint{APS/123-QED}

\title{Possible Experimental Realization of a Basic \textit{Z}$_\mathbf{2}$ Topological Semimetal}% Force line breaks with \\

\author{Erik Haubold}
\author{Alexander Fedorov}
\affiliation{IFW Dresden, Helmholtzstr. 20, 01069 Dresden, Germany }%

\author{Igor P. Rusinov}
\affiliation{Tomsk State University, pr. Lenina 36, 634050 Tomsk, Russia}
\affiliation{St. Petersburg State University, Universitetskaya nab. 7/9, 199034 St. Petersburg, Russia}
\author{Tatiana V. Menshchikova}
\affiliation{Tomsk State University, pr. Lenina 36, 634050 Tomsk, Russia}

\author{Viola Duppel}
\affiliation{Max Planck Institute for Solid State Research, Heisenbergstr. 1, 70569 Stuttgart, Germany}
\author{Daniel Friedrich}
\affiliation{Institut f{\"u}r Anorganische Chemie, Universit{\"a}t Regensburg, 93040 Regensburg}
\author{Florian Pielnhofer}
\affiliation{Max Planck Institute for Solid State Research, Heisenbergstr. 1, 70569 Stuttgart, Germany}
\affiliation{Institut f{\"u}r Anorganische Chemie, Universit{\"a}t Regensburg, 93040 Regensburg}

\author{Richard Weihrich}
\affiliation{Universität Augsburg, Institut f{\"u}r Materials Ressource Management, Universit{\"a}tsstr. 2, 86135 Augsburg }

\author{Arno Pfitzner}
\affiliation{Institut f{\"u}r Anorganische Chemie, Universit{\"a}t Regensburg, 93040 Regensburg}

\author{Alexander Zeugner}
\author{Anna Isaeva}
\affiliation{Faculty of Chemistry and Food Chemistry, Technische Universit{\"a}t Dresden, Helmholtzstr. 10, 01069 Dresden, Germany}

\author{Setti Thirupathaiah}
\altaffiliation[present address: ]{S. N. Bose National Centre for Basic Sciences, Block-JD, Sector-III, Salt Lake, Kolkata 700106, India}
\affiliation{IFW Dresden, Helmholtzstr. 20, 01069 Dresden, Germany }%

\author{Yevhen Kushnirenko}
\affiliation{IFW Dresden, Helmholtzstr. 20, 01069 Dresden, Germany }%

\author{Emile Rienks}
\affiliation{IFW Dresden, Helmholtzstr. 20, 01069 Dresden, Germany }%
\affiliation{Department of Physics, TU Dresden, 01062 Dresden, Germany}

\author{Timur Kim}
\affiliation{ Diamond Light Source, Harwell Campus, Didcot OX11 0DE, United Kingdom }

\author{Evgueni V. Chulkov}
\affiliation{Tomsk State University, pr. Lenina 36, 634050 Tomsk, Russia}
\affiliation{St. Petersburg State University, Universitetskaya nab. 7/9, 199034 St. Petersburg, Russia}
\affiliation{Donostia International Physics Center, Paseo de Manuel Lardizabal 4, 20018 San Sebastian/Donostia, Basque Country, Spain}
\affiliation{Departamento de Fisica de Materiales, Facultad de Ciencias Quimicas, and Centro de Fisica de Materiales and Materials Physics Center, University of the Basque Country (UPV/EHU), 20080 San Sebastian/Donostia, Basque Country, Spain}

\author{Bernd B{\"u}chner}
\affiliation{IFW Dresden, Helmholtzstr. 20, 01069 Dresden, Germany }%
\affiliation{Department of Physics, TU Dresden, 01062 Dresden, Germany}

\author{Sergey Borisenko}
\affiliation{IFW Dresden, Helmholtzstr. 20, 01069 Dresden, Germany }%


%\\collaboration{MUSO Collaboration}%\noaffiliation


%\\collaboration{CLEO Collaboration}%\noaffiliation

\date{\today}% It is always \today, today,

\begin{abstract}

We report experimental and theoretical evidence that \ch{GaGeTe} is a basic $Z_2$ topological semimetal with three types of charge carriers: bulk-originated electrons and holes as well as surface state electrons. This electronic situation is qualitatively similar to the primer 3D topological insulator~\ch{Bi2Se3}, but important differences account for an unprecedented transport scenario in~\ch{GaGeTe}. High-resolution angle-resolved photoemission spectroscopy combined with advanced band structure calculations show a small indirect energy gap caused by a peculiar band inversion in the \textit{T}-point of the Brillouin zone in~\ch{GaGeTe}. An energy overlap of the valence and conduction bands brings both electron- and hole-like carriers to the Fermi level, while the momentum gap between the corresponding dispersions remains finite. We argue that peculiarities of the electronic spectrum of \ch{GaGeTe} have a fundamental importance for the physics of topological matter and may boost the material's application potential.

\end{abstract}

\maketitle

%\tableofcontents

A variety of materials where topology of the electronic structure plays a special role in transport properties is rapidly growing \cite{ando2013, ando2015, armitage2018}. The envisioned applications of topological materials in  novel devices and quantum information technology will be strongly influenced by the fine balance between their charge carriers of various types. In classic 3D topological insulators, like \ch{Bi2Se3}, the transport properties are dictated by the non-degenerate massless topological surface states as well as the electron-like carriers from the bulk conduction band~\cite{hasan2010}. In 3D Dirac or Weyl semimetals, the Fermi surface is formed by single points of band degeneration (type I), or hole and electronic pockets touching in discrete points (type II) \cite{armitage2018}. %The latter are non-degenerate in Weyl semimetals and so are the Fermi surface arcs, which also contribute to the exotic transport properties.
A bulk material with a gapped electronic spectrum combining a comparable number of both types of the bulk charge carriers with the spin-momentum locked topological surface states has not yet been accounted for. Here we put forward \ch{GaGeTe} as the first example of such conceptually different electronic situation, i.~e. a basic $Z_2$ topological semimetal.

%TODOComment: shall we mention zero-gap semiconductors like HgTe, K3Bi as yet another example of a different situation?

The layered compound \ch{GaGeTe} was first synthesized and structurally characterized quite long ago %(G. Kra, R. Eholie, J. Flahant, Comples rendus des seances de l ́academie des sciences, Serie C: Science Chimiques 284 (1977) 889–892.)
\cite{kra1977, fenske1983}, and its phonon structure %(E. Lopez-Cruz, PRB 1984 Raman spectrum and lattice dynamics of GaGeTe)
\cite{lcruz1984}, thermoelectric %(AIP Conference Proceedings 1449, 267 (2012); https://doi.org/10.1063/1.4731548)
\cite{aipconference2012} and transport %(Journal of Crystal Growth 380 (2013) 72–77)
\cite{kucek2013} properties were consequently studied. However, the first theoretical study of its bulk electronic structure has appeared very recently 
\cite{GGT_theory} and has sparked strong interest in the surface electronic structure of this material \cite{zhang2017}. %(Zhang et al  J. Mater. Chem. C, 2017, 5, 8847).
In \cite{GGT_theory} structural and electronic resemblance between a structure fragment of the layered \ch{GaGeTe} bulk and a heavy analogue of graphene, germanene \cite{davila2014}, has been highlighted. 2D materials with buckled honeycomb atomic arrangements, e.g. silicene, germanene, stanene, are promising for the realization of new devices. For instance, silicene demonstrates such advantages as high carrier mobility, excellent mechanical flexibility and compatibility with existing Si-based electronics. Most recent studies establish theoretically \cite{zhang2017} %(Zhang et al  J. Mater. Chem. C, 2017, 5, 8847)
 and experimentally %(APPLIED PHYSICS LETTERS 111, 203504 (2017) Ultrathin GaGeTe p-type transistors, Weike Wang)
\cite{wang2017} that monolayers of \ch{GaGeTe}, namely six-atom-thick sheets (Fig. 1a) held together by van der Waals interaction, can be exfoliated and are also suitable for the fabrication of nanodevices including transistors and photodetectors. Ultrathin films of \ch{GaGeTe} exhibit transport characteristics superior to many other FETs based on 2D-materials, such high hole mobility and good on/off current ratios.

\ch{GaGeTe} is not only promising for potential semiconductor applications, but is also intriguing from the fundamental perspective. First-principle calculations performed by co-authors in Ref. \cite{GGT_theory} have shown a possibility of a topological band inversion in its electronic spectrum driven by spin-orbit coupling. The valence and the conduction bands invert at the \textit{T}-point, opening up a narrow indirect gap of the order of \SI{30}{\milli\electronvolt}. The bandgap size appeared to be very sensitive to the chosen computational parameters. Earlier published experimental studies of physical properties of \ch{GaGeTe} neither corroborate nor refute this scenario. For instance, an optical study implies that \ch{GaGeTe} is a semiconductor with the band gap of \SI{1.12}{\electronvolt} \cite{kucek2013}. %(Journal of Crystal Growth 380 (2013) 72–77). 
To shed more light on these controversial outcomes of theory and experiment, we have engaged on the present study of high-quality \ch{GaGeTe} single crystals by synchrotron-based angle-resolved photoemission spectroscopy (ARPES) with variable photon energies. We clarify the details of the electronic structure of \ch{GaGeTe} and scrutinize its probable topological nature with an aid of state-of-the-art calculations of its bulk and surface electronic structures. 

%Z2 stuff

\begin{figure}[bt]
    \centering
    \includegraphics[width=\linewidth]{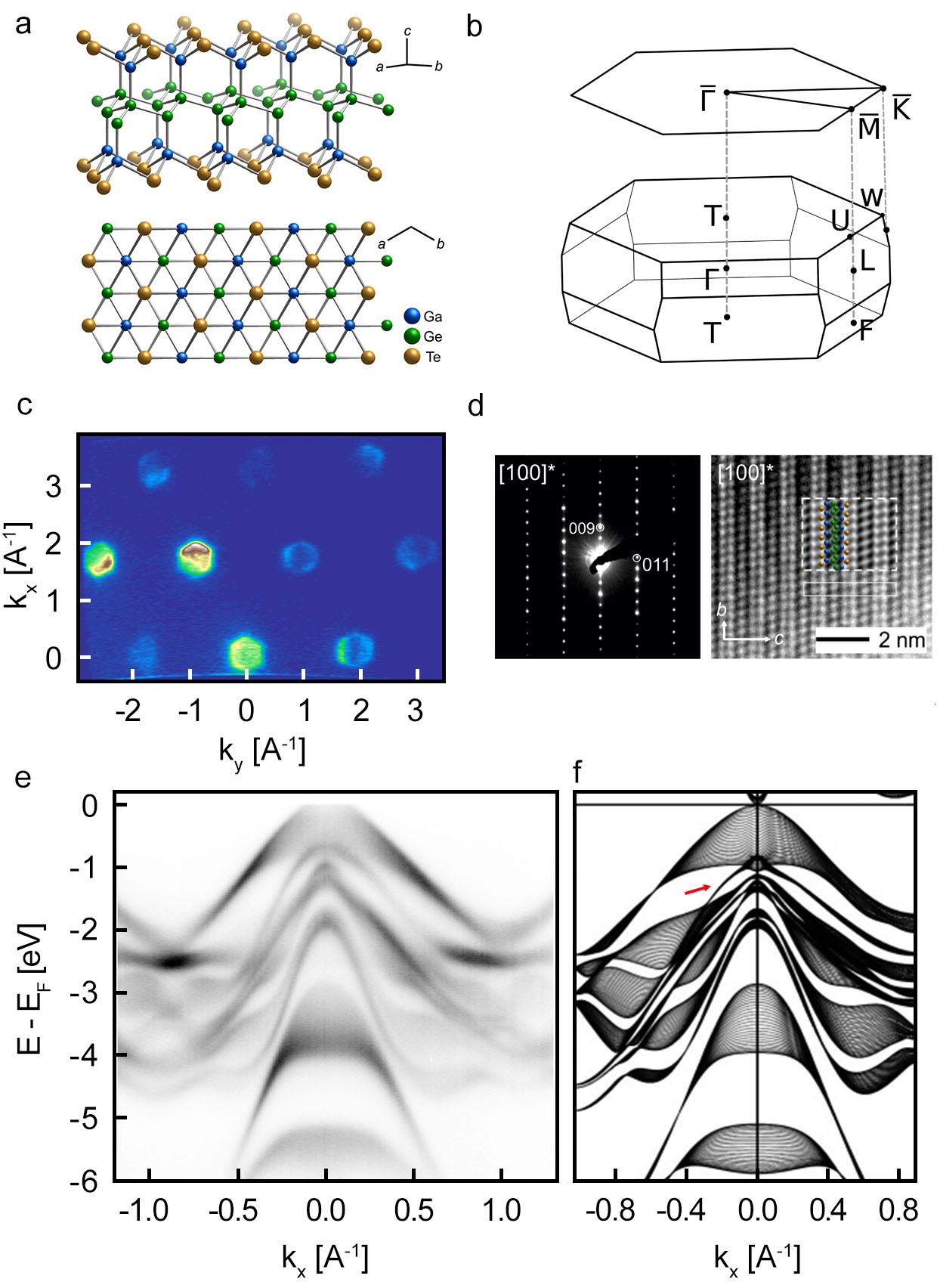}%d
    \caption{a: One sixtuple layer of \ch{GaGeTe} (side and top view), the building unit of the bulk layered structure. The germanene-like buckled Ge-bilayer is highlighted in green. b: Trigonal 3D Brillouin zone and its projection onto the 2D BZ. c: Fermi surface overview map taken at \SI{180}{\electronvolt}. d: HRTEM image e: Energy momentum cut taken at \SI{100}{\electronvolt}. f: calculation of $\overline{\mathit{K}}-\overline{\mathit{\Gamma}}-\overline{\mathit{M}}$. The influence of the asymmetric Brillouin zone is clearly seen. This bandstructure shows bands for multiple k$_\text{z}$ values as multiple lines. The only major difference to e) is highlighted by a red arrow.} %better formulate with band folding?
    \label{fig:fig1}
\end{figure}

%\textcolor{red}{
%The crystal structure of \ch{GaGeTe} is shown in Fig. \ref{fig:fig1}a and corresponding Brillouin zone (BZ) in Fig. 1b. This is a layered system with a hexagonal unit cell with the space group $R\overline{3}m$ (Fig (\ref{fig:fig1})). One layer of the material consists of two Ga-Te sheets with a Ge sheet in between. The surface termination is always Te, which is important for photoemission measurements.
%}

%\textcolor{blue}{
\ch{GaGeTe} crystallizes in a trigonal unit cell (sp. gr. $R\overline{3}m$) that comprises an $ABC$-stacking of sixtuple layers. High-resolution transmission electron microscopy experiments confirm the ordering along the stacking direction (Fig. \ref{fig:fig1}d). One building unit of \ch{GaGeTe} (Fig. \ref{fig:fig1}a) can be regarded as a corrugated Ge-bilayer (germanene) sandwiched between two \ch{GaTe} sheets of the zincblende type. Bonding within the sixtuple layers is essentially covalent, while the inter-layer interaction is of the van der Waals type. Consequently, the natural cleavage plane of \ch{GaGeTe} crystals is always Te-terminated, which facilitates the interpretation of the photoemission measurements. 
Cleavage in ultra-high vacuum results in a mirror-like surface which allows to record, in particular, large Fermi surface maps.

One of such FS maps taken using the \SI{180}{\electronvolt} horizontally polarized photons is shown in Fig. \ref{fig:fig1}c. The FS contour is very small, having the radius of \SI{0.35}{\per\angstrom}, it is however larger than that of \ch{Bi2Se3} \cite{kordyuk2011}. %\textcolor[rgb]{0,0,1}{larger, but depending a bit on crystal(doping) (100meV)}.
The size, the shape and intensity of this contour depends on momentum and thus imply three-dimensionality of the electronic structure: effective k$_\text{z}$ probed by photons with particular energy is always smaller for larger k$_\text{x}$ and k$_\text{y}$. For example, the signal is nearly absent for the contour at (\SI{0}{\per\angstrom}, \SI{3.5}{\per\angstrom}) and its size is smaller for the contour centered at (\SI{-1.8}{\per\angstrom}, \SI{0}{\per\angstrom}). It means that electron escape depth in this material is sufficient to distinguish between different k$_\text{z}$'s even using the same photon energy.

We compare the underlying dispersions, experimental and calculated ones, in Fig. \ref{fig:fig1}d and e, respectively. As a first step it is instructive to compare with the k$_\text{z}$-integrated calculations to find out about the k$_\text{z}$-resolution mentioned above. Experimental data are taken along the cut running parallel to the $\overline{\mathit{K}}-\overline{\mathit{\Gamma}}-\overline{\mathit{M}}$ direction of the 2D Brillouin zone. The projection of the 3D BZ onto the 2D BZ is shown in Fig. \ref{fig:fig1}b. The overall agreement is remarkable and not only qualitative, especially for the $\overline{\mathit{K}}-\overline{\mathit{\Gamma}}$ direction: all the dispersions coincide, including the energy positions of the dispersion maxima at zero momentum. The only exception is the two-dimensional band (weak k$_\text{z}$-dispersion) having a top at $\sim$\SI{0.8}{\electronvolt} according to the calculations while experiment shows it reaches the maximum at $\sim$\SI{0.65}{\electronvolt} binding energy (see the red arrow in Fig. \ref{fig:fig1}e). One can also notice that the experiment has more spectral weight at particular k$_\text{z}$ values, although an admixture from other k$_\text{z}$'s is still visible. This observation agrees well with the moderate k$_\text{z}$-resolution concluded from the Fermi surface map shown in panel c. Within the considered energy range, we note a good agreement between experiment and theory, independent of which exchange-correlation functional, PBE or HSE, is used to describe the occupied electronic structure of \ch{GaGeTe} \cite{GGT_theory}.

%%BEGIN FIG3
\begin{figure*}[tb]
    \centering
    \includegraphics[width=\linewidth]{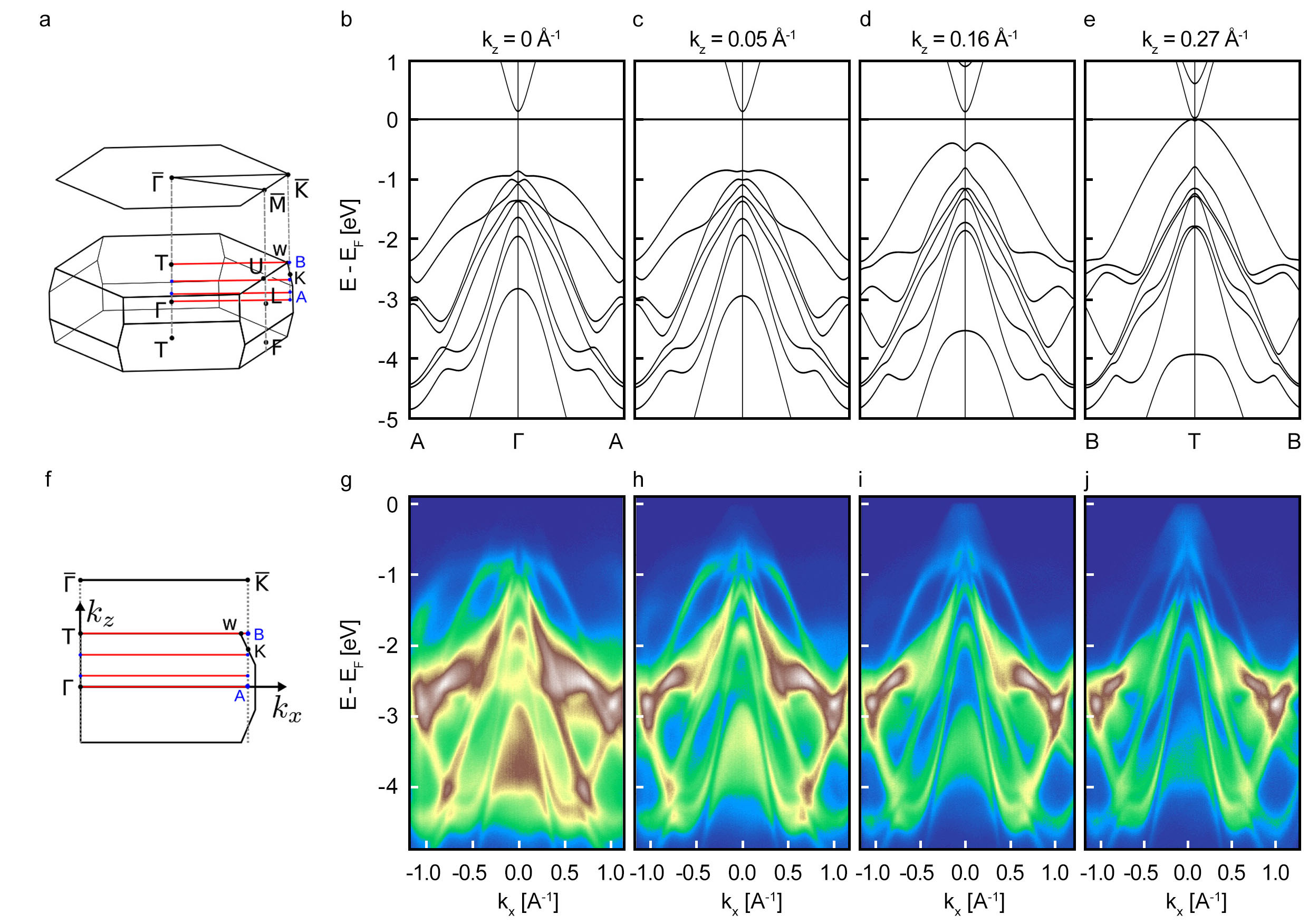}%f
    %\caption{k$_\text{z}$ behaviour of the electronic structure. a-d: calculated bandstructure from $\mathit{\Gamma}$ (a) to \textit{T} (d). e-h: Energy momentum cuts taken with \SI{91}{\electronvolt} to \SI{100}{\electronvolt} in equidistant steps, showing good agreement with the calculations. }
    \caption{Bulk Brillouin zone (a, f) with k$_\text{z}$ constant lines (red), along which the electronic spectra (b-e) were calculated, corresponding to the electronic band structure in the order of increasing k$_\text{z}$. (g-j) Energy momentum cuts taken with 91 eV to 100 eV in equidistant steps. }
    \label{fig:fig3}
\end{figure*}

Now we can address its k$_\text{z}$ dependence in more details by comparing the data collected with differing photon energies. The typical data representing the k$_\text{z}$ behaviour are shown in Fig. \ref{fig:fig3} and compared with the results of the band structure calculations. Panels a-d show a calculated band structure using the PBE functional. The band inversion manifests itself very clearly in the cuts running through the \textit{T}-point (Fig. \ref{fig:fig3}d). At other k$_\text{z}$'s one can still notice the characteristic dip in the quickly dispersing hole-like band. This trend is confirmed by the experimental data shown in panels e-h. These data were measured using the photon energies from \SI{91}{\electronvolt} (e) to \SI{100}{\electronvolt} (h) in steps of \SI{3}{\electronvolt}. The hole-like dispersion is noticeably changing when approaching the Fermi level. In addition to this, one can see that another feature is present at the Fermi level and its intensity is growing together with the intensity of the hole-like feature. There is no energy gap visible between them. At higher binding energies there are bands showing a large k$_\text{z}$ dependency as well, e.g. the bands seen in Fig. \ref{fig:fig1}e,f at \SIlist{-1;-3;-6}{\electronvolt} at the $\mathit{\Gamma}$ point. The corresponding intensity redistribution in the experimental data shown in panels e-h of Fig. \ref{fig:fig3} clearly tracks these changes.  As expected, the bands with weaker k$_\text{z}$ dispersion are visible in all energy momentum cuts at approximately the same energies.

Summarizing the experimental observations so far, we can conclude that the energy gap is either very small or absent at all, thus, annulling the earlier beliefs that \ch{GaGeTe} is a wide-gap semiconductor with a \SI{1.1}{\electronvolt} gap. This, together with the peculiar shape of the closest to E$_\text{F}$ dispersion of the hole-like band clearly speaks in favour of the band inversion and, subsequently, the topological nature of \ch{GaGeTe}. In this case, the topological surface states are expected.

%%BEGIN FIG2
\begin{figure}[tb]
    \centering
    \includegraphics[width=\linewidth]{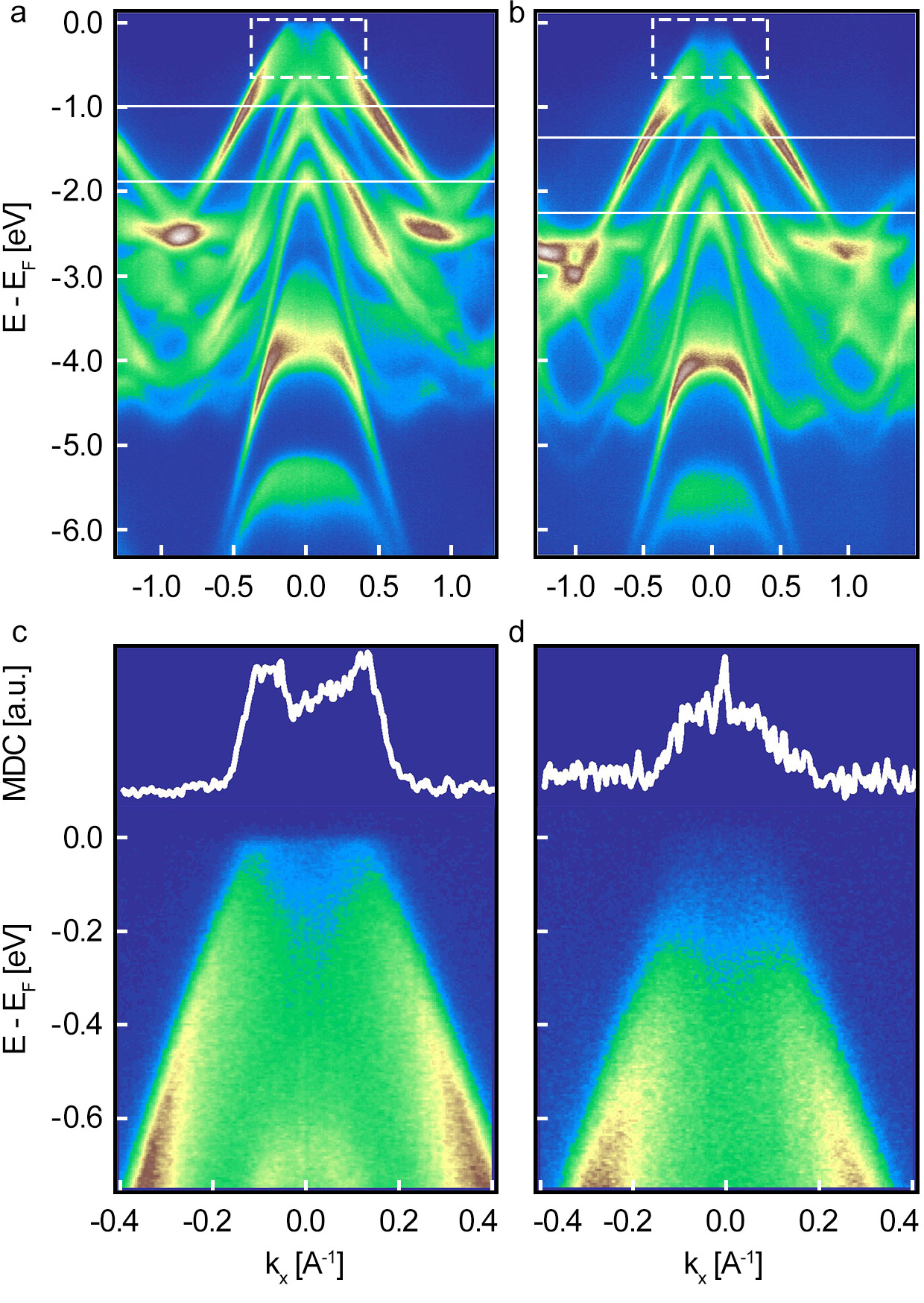}%fig2g
    \caption{Results of the dosing of single crystals with potassium. a: before dosing, b: after dosing approximately half a monolayer. The center of one distinct band is highlighted with a white line. c and d: zoomed-in pictures close to the Fermi energy, with their respective locations shown with dashed boxes in the respective panels above. The outline of the dispersion is highlighted by a dashed line, and the momentum distribution curve of the Fermi energy is shown above the data.}
    \label{fig:fig2}
\end{figure}

To clarify whether the PBE-based result, namely the topological band inversion, holds in terms of the gap size, we have carried out typical dosing experiments evaporating potassium on the surface of the sample. The results of these measurements are shown in Fig \ref{fig:fig2}. The dosing turned out to be effective, and we managed to shift the chemical potential by nearly half an eV (Fig. \ref{fig:fig2}a,b). As it was suggested by Fig. \ref{fig:fig1}d,e, the top of the hole-like band was very close to the Fermi level (Fig. \ref{fig:fig2}c,d). According to DFT calculations, in the case of the HSE functional, the band structure is topologically trivial with a rather large direct gap (\SI{550}{\milli\electronvolt}) with the gap extrema located exactly in the \textit{T}-point of 3D BZ \cite{GGT_theory}. In the case of the PBE functional, the band gap size is smaller (∼\SI{30}{\milli\electronvolt}) and the gap is indirect, implying the  topological character of the electronic spectrum.

In accordance with the PBE-based calculations, we have not observed a large energy gap suggested by other functionals including HSE either. On the other hand, we also have not observed a small direct energy gap between the valence and conduction bands and a sharp electron-like bottom of the conduction band itself, suggested by the PBE calculations. Instead, we detected a spectral density at the Fermi level (Fig. \ref{fig:fig2}d) with the finite momentum width, which stayed nearly constant and increased only slightly up to the highest dosing level. At higher rates the features became too blurred to encourage further dosing. We found out that this intensity could be enhanced using the light of different photon energies, implying that there is indeed no direct gap between the valence and conduction bands and one can safely rule out the scenario with a large gap. The electronic structure of \ch{GaGeTe} is thus characterized by a band inversion and is non-trivial.

We have analysed the topological phase in \ch{GaGeTe} by two methods. At first, the parities of the states at the time-reversal invariant momentum (TRIM) points of the primitive rhombohedral cell were considered. The results establish a non-trivial character of the electronic structure, as characterized by the topological invariant $Z_2=1;001$, which originates from a bulk band inversion at the \textit{T}-point of the 3D Brillouin zone. The edge of the valence zone is constituted by the even states, the s-orbitals of the Ge atoms, whereas the conduction-band edge is composed of the odd states, the Te p-orbitals.  Away from the region of the band inversion, the orbital composition of the gap edges is reversed.  We also checked the topological nature of \ch{GaGeTe} by varying spin-orbit coupling strength. The gap size decreases down to zero when the spin-orbit coupling strength $\lambda/\lambda_0$ is diminished to $\sim 0.65$, where $\lambda_0$ is the natural value of the SOC contribution. At smaller $\lambda/\lambda_0$, the electronic spectrum is trivial, the gap edges are not inverted, and $Z_2$ is equal to 0;(000).  
Additionally, we also prove the non-trivial topology by the Wilson loop (WL) method proposed in \cite{soluyanov2011, yu2011}.  It allows to trace the topology of a material graphically. The WL spectra at $k_z=0$ and $k_z=0$ are shown in  supplementary Data. %Fig. \ref{fig:suppl1}. 
It is evident that the spectra cross an arbitrary reference line an odd number of times at $k_z=0.5$ and an even number of times at $k_z=0$. This case corresponds to $Z_2$=1;(001). Also at $\lambda/\lambda_0<0.65$, both for $k_z=0.5$ and  $k_z=0$ the WL bands cross the reference line an even number of times. This finding is also in agreement with the above given analysis of the $Z_2$ invariant conducted on the basis of the parity of the wave-functions.

%BEGIN FIG4
\begin{figure}[tb]
    \centering
    \includegraphics[width=\linewidth]{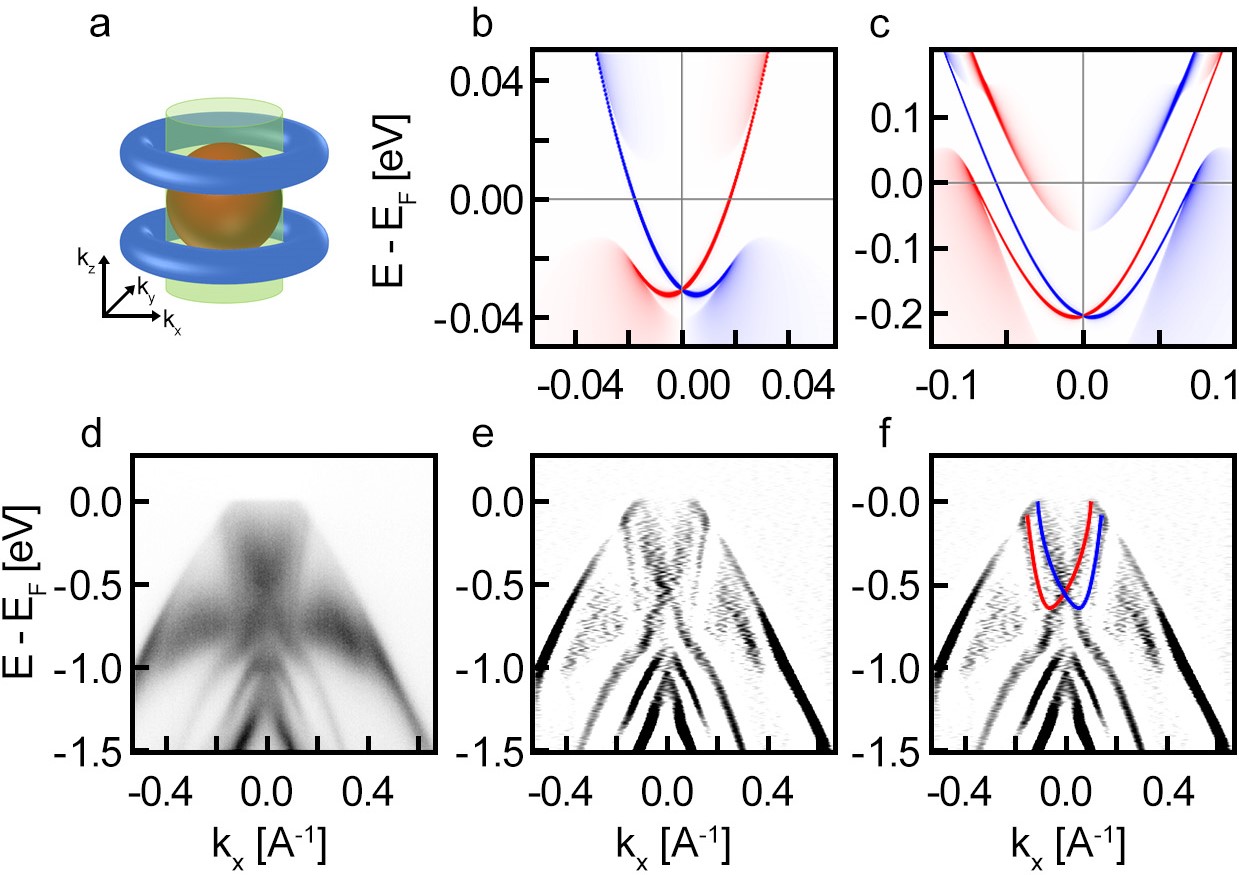}%f
    \caption{a: Schematic of the Fermi surface with an electron pocket (red), hole pockets (blue) and surface states (green).
    b,c: Spin-resolved surface spectral function calculated within {\it ab initio} based tight binding model with the experimentally determined unit cell parameters (b) and with the unit cell parameter $a$ expanded by 1 \% (c). Spin-polarisation is color-coded (red and blue). Spectrum (b) shows the initially expected direct band gap, whereas spectrum (c) highlights an indirect band gap actually found by photoemission experiments.
    %b, c: calculation at smaller area with different parameters. Highlighting the originally expected direct gap (b) and the actually measured indirect gap (c). The colors correspond to the spin-polarization. 
    d: \SI{19}{\electronvolt} energy momentum cut taken at the T-point. e: 2$^\text{nd}$ derivative of d, highlighting the edges in panel d. f: same as e, but with the dispersions identified as surface states marked in red and blue.}
    \label{fig:fig4}
\end{figure}

In Fig. \ref{fig:fig4} we present the data taken using lower photon energies, \SI{19}{\electronvolt}, to zoom in the region of interest. As we noticed earlier, it is difficult to assign the particular high-symmetry point along the $\mathit{\Gamma}$--\textit{T} pathway to a particular photon energy because of the moderate k$_\text{z}$-resolution in the case of \ch{GaGeTe}. Also in this case, on one side, the higher intensity of the strongly dispersing hole-like band at the higher binding energies (approx. \SI{0.75}{\electronvolt}) would imply this cut corresponds to the vicinity of the $\mathit{\Gamma}$-point; on the other side, the strongest spectral weight at the Fermi level means one is close to the \textit{T}-point. In any case, now it is clearly seen that the calculations with the experimental lattice parameters do not reproduce the photoemission experiment in full details. The most striking observation is that the electron-like dispersion corresponding to the bottom of the conduction band is clearly seen below the Fermi level. We note that this is not a signature of the n-doping since Figures \ref{fig:fig1}-\ref{fig:fig3} demonstrated that also the valence hole-like band crosses the chemical potential. Taking into account the overall agreement with the calculations on a larger energy scale, we conclude that the deviations preserving a Luttinger count occur. Instead of a semiconductor with a tiny direct gap, experiment shows that \ch{GaGeTe} is a semimetal with the small bulk-originated Fermi surfaces of both types, a hole-like FS, with the shape of a torus due to the crossings of the strongly dispersing hole-like band, and an electron-like FS, with the shape of an ellipsoid or even a sphere, supported by the electron-like conduction band with its minimum below the Fermi level near the \textit{T}-point. Our calculations demonstrate that the band dispersions in the vicinity of the Fermi energy are extremely sensitive to even miniscule changes in the unit cell parameters and geometry optimization. For instance, the described experimental picture can be confirmed computationally with a slightly extended unit cell parameter $a$, namely, from \SI{4.08}{\angstrom} to \SI{4.13}{\angstrom}. To keep the cell volume constant, the lattice parameter $c$ has been contracted from \SI{34.54}{\angstrom} to \SI{33.90}{\angstrom}. In both cases, the material is topological and qualitatively similar surface states should be present. We identify those as straight linear dispersions accompanying the conduction band states down to approx. \SI{0.6}{\electronvolt} in the experimental spectra shown in panel d. Second derivative (panel e) helps to distinguish them from the bulk projections of the valence and conduction bands.

Although the situation in \ch{GaGeTe} is qualitatively similar to the simplest 3D topological insulator \ch{Bi2Se3}, there are important differences. First is that \ch{GaGeTe} does not have a direct gap and is a semimetal in contrast to \ch{Bi2Se3}\cite{nechaev2013}, i.e. the valence and the conduction bands overlap in energy and both cross the Fermi level resulting in small Fermi surfaces. Second, the sizes of an energy gap at a particular k-point and of a momentum gap at particular energy are much smaller: less than \SI{200}{\milli\electronvolt} and less than \SI{0.1}{\per\angstrom} respectively, which is in agreement with the calculations shown in Fig. \ref{fig:fig4}. The surface states themselves are less robust than in \ch{Bi2Se3}. This is related to the quality of the surface after the cleavage of the sample in ultra-high vacuum. In spite of the layered structure of \ch{GaGeTe}, it is very difficult to obtain a relatively big shiny portion of the atomically clean surface. In such cases the surface states are known to be elusive and require repeating experiments to be detected. One of such successful attempts is documented in Fig. \ref{fig:fig4}. The surface states support another small Fermi surface, which we schematically depict in panel (a) as a cylinder together with the bulk originated 3D Fermi surfaces. Our calculations show that these topological surface states are spin-polarized.

The bulk \ch{GaGeTe} thus emerges as a basic $Z_2$ topological semimetal. There is only one Dirac cone formed by the topological surface states per Brillouin zone. Unlike the similar electronic structure of Sb-Bi-Sb heterostructures \cite{takayama2014, bihlmayer2010}, the number of bulk charge carriers of the opposite types is the same in \ch{GaGeTe}. As previous \cite{GGT_theory, zhang2017, wang2017} and this studies have demonstrated, the electronic structure of \ch{GaGeTe} is very sensitive to a number of parameters and therefore is very attractive for nanodevice fabrication. It can be tuned by strain, exfoliation, doping and gating. \ch{GaGeTe} is stable on air, not moisture-senstitive, much more ecologically friendly than compounds containing heavy bismuth. At the same time the monolayer of \ch{GaGeTe} seems to be dynamically and thermodynamically stable at very high temperatures \cite{zhang2017}. Having such a peculiar Fermi surface (Fig. \ref{fig:fig4} a) with two 3D sheets and one 2D sheet, the hypothetical shift of the Fermi level can switch the system from having purely bulk n-type charge carriers to having purely p-type ones. In between these two extremes, the presence of the non-degenerate topological surface states with the spin-texture can be exploited.

The individual features of this material thus boost its application potential and have a fundamental importance for the physics of topological matter.

This work was supported under DFG grant BO 1912/7-1, IS 250/2-1 of the SPP 1666 program, RU 776/15-1 of the ERANET-Chemistry program, Tomsk State University Project 8.1.01.2018 and St. Petersburg University Project 15.61.202.2015. The authors acknowledge Diamond Light Source for the beamtime at I05 beamline under proposal SI18586 as well as the BESSY II Berlin for the beamtime at 1$^\text{3}$ ARPES station under proposals 171-05051CR and 172-05659CR/R.

\bibliography{lit}
%TODO: build standalone supple
%\appendix
%\clearpage
%\input{actual_suppl}
\end{document}